\documentclass[
  twocolumn,
  showpacs,
  preprintnumbers,
  amsmath,
  amssymb,
  aps,
  prc,
  floatfix,
  letterpaper,
  superscriptaddress
]{revtex4-1}

\usepackage{tabularx}
\usepackage{graphicx}
\usepackage{dcolumn}     
\usepackage{bm}          

\def\today{\ifcase\month\or
  January\or February\or March\or April\or May\or June\or
  July\or August\or September\or October\or November\or December\fi
  \space\number\day, \number\year}

\newcommand\NSU{Norfolk State University, Norfolk, Virginia 23504}
\newcommand\UNH{University of New Hampshire, Durham, New Hampshire 03824}
\newcommand\UVa{University of Virginia, Charlottesville, Virginia 22903}
\newcommand\Jlab{Thomas Jefferson National Accelerator Facility,
  Newport News, Virginia 23606 }
\newcommand\UBasel{Universit\"{a}t Basel, CH-4056 Basel, Switzerland}
\newcommand\FIU{Florida International University, Miami, Florida 33199}
\newcommand\HamptonU{Hampton University, Hampton, Virginia 23668}
\newcommand\MissSU{Mississippi State University, Mississippi State,
  Mississippi 39762}
\newcommand\NCAT{North Carolina A\&T State University, Greensboro,
  North Carolina 27411}
\newcommand\ODU{Old Dominion University, Norfolk, Virginia 23529}
\newcommand\SUNO{Southern University at New Orleans, New Orleans,
  Louisiana 70126}
\newcommand\TelAviv{Tel Aviv University, Tel Aviv, 69978 Israel}
\newcommand\UMD{University of Maryland, College Park, Maryland 20742}
\newcommand\UNCW{University of North Carolina, Wilmington, North Carolina 28403}
\newcommand\VaTech{Virginia Polytechnic Institute \& State University,
Blacksburg, Virginia 24061}
\newcommand\Yerevan{Yerevan Physics Institute, Yerevan, Armenia 0036}
\newcommand\Kyungpook{Kyungpook National University, Daegu 702-701, South Korea}
\newcommand\UMass{University of Massachusetts,Amherst, Massachusetts 01003}
\newcommand\PSI{Paul Scherrer Institut, Villigen, Switzerland}

\begin{document}

\title{Probing Quark-Gluon Interactions with Transverse Polarized Scattering}

\author{K.~Slifer}     \affiliation{\UVa}\affiliation{\UNH}
\author{O.A.~Rond\'{o}n}        \affiliation{\UVa}
\author{A.~Aghalaryan}      \affiliation{\Yerevan}
\author{A.~Ahmidouch}        \affiliation{\NCAT}
\author{R.~Asaturyan}        \affiliation{\Yerevan}
\author{F.~Bloch}        \affiliation{\UBasel}
\author{W.~Boeglin}        \affiliation{\FIU}
\author{P.~Bosted}        \affiliation{\Jlab}
\author{C.~Carasco}        \affiliation{\UBasel}
\author{R.~Carlini}        \affiliation{\Jlab}
\author{J.~Cha}         \affiliation{\MissSU}
\author{J.P.~Chen}        \affiliation{\Jlab}
\author{M.E.~Christy}        \affiliation{\HamptonU}
\author{L.~Cole}        \affiliation{\HamptonU}
\author{L.~Coman}        \affiliation{\FIU}
\author{D.~Crabb}        \affiliation{\UVa}
\author{S.~Danagoulian}     \affiliation{\NCAT}
\author{D.~Day}         \affiliation{\UVa}
\author{J.~Dunne}        \affiliation{\MissSU}
\author{M.~Elaasar}        \affiliation{\SUNO}
\author{R.~Ent}         \affiliation{\Jlab}
\author{H.~Fenker}        \affiliation{\Jlab}
\author{E.~Frlez}        \affiliation{\UVa}
\author{D.~Gaskell}                \affiliation{\Jlab}
\author{L.~Gan}                \affiliation{\UNCW}
\author{J.~Gomez}        \affiliation{\Jlab}
\author{B.~Hu}          \affiliation{\HamptonU}
\author{J.~Jourdan}        \affiliation{\UBasel}
\author{M.~K.~Jones}        \affiliation{\Jlab}
\author{C.~Keith}        \affiliation{\Jlab}
\author{C.E.~Keppel}        \affiliation{\HamptonU}
\author{M.~Khandaker}        \affiliation{\NSU}
\author{A.~Klein}        \affiliation{\ODU}
\author{L.~Kramer}        \affiliation{\FIU}
\author{Y.~Liang}        \affiliation{\HamptonU}
\author{J.~Lichtenstadt}    \affiliation{\TelAviv}
\author{R.~Lindgren}        \affiliation{\UVa}
\author{D.~Mack}        \affiliation{\Jlab}
\author{P.~McKee}        \affiliation{\UVa}
\author{D.~McNulty}         \affiliation{\UVa}\affiliation{\UMass}
\author{D.~Meekins}        \affiliation{\Jlab}
\author{H.~Mkrtchyan}        \affiliation{\Yerevan}
\author{R.~Nasseripour}     \affiliation{\FIU}
\author{I.~Niculescu}        \affiliation{\Jlab}
\author{K.~Normand}        \affiliation{\UBasel}
\author{B.~Norum}        \affiliation{\UVa}
\author{D.~Pocanic}        \affiliation{\UVa}
\author{Y.~Prok}        \affiliation{\UVa}
\author{B.~Raue}        \affiliation{\FIU}
\author{J.~Reinhold}        \affiliation{\FIU}
\author{J.~Roche}        \affiliation{\Jlab}
\author{D.~Kiselev (nee Rohe)}        \affiliation{\UBasel}\affiliation{\PSI}
\author{N.~Savvinov}        \affiliation{\UMD}
\author{B.~Sawatzky}        \affiliation{\UVa}
\author{M.~Seely}        \affiliation{\Jlab}
\author{I.~Sick}        \affiliation{\UBasel}
\author{C.~Smith}        \affiliation{\UVa}
\author{G.~Smith}        \affiliation{\Jlab}
\author{S.~Stepanyan}    \affiliation{\Kyungpook}
\author{L.~Tang}        \affiliation{\HamptonU}
\author{S.~Tajima}         \affiliation{\UVa}
\author{G.~Testa}        \affiliation{\UBasel}
\author{W.~Vulcan}        \affiliation{\Jlab}
\author{K.~Wang}        \affiliation{\UVa}
\author{G.~Warren}        \affiliation{\UBasel}\affiliation{\Jlab}
\author{F.R.~Wesselmann}    \affiliation{\UVa}\affiliation{\NSU}
\author{S.~Wood}        \affiliation{\Jlab}
\author{C.~Yan}         \affiliation{\Jlab}
\author{L.~Yuan}        \affiliation{\HamptonU}
\author{J.~Yun}         \affiliation{\VaTech}
\author{M.~Zeier}        \affiliation{\UVa}
\author{H.~Zhu}         \affiliation{\UVa}

\collaboration{The Resonance Spin Structure Collaboration}
\noaffiliation

\date{\today}

\begin{abstract}
We have extracted QCD matrix elements from our data on doubly polarized
inelastic scattering of electrons on nuclei. We find the higher twist
matrix element $\tilde{d_2}$, which arises strictly from quark-gluon
interactions, to be unambiguously non zero. The data also reveal an
isospin dependence of higher twist effects 
if we assume that the Burkhardt-Cottingham Sum rule is valid.
The fundamental  Bjorken sum rule
obtained from the $a_0$ matrix element is satisfied at our low momentum
transfer. 
\end{abstract}


\maketitle


Polarized deep inelastic scattering (DIS) of leptons 
on nucleons has helped to establish the modern understanding of nucleon structure. 
Measurements of the longitudinal nucleon spin structure function (SSF) $g_1$ 
in inclusive doubly polarized DIS, where only the final lepton is detected, 
have revealed that the nucleon spin does not result from the naive sum of the 
spins of its three valence quarks. Rather, contributions from sea quarks, 
gluon spin and orbital angular momentum seem to be
required~\cite{Abe:1998wq,*Anthony:2000fn,*Anthony:2002hy,Airapetian:2007mh,Amarian:2003jy}.
DIS is well understood at  high energies, when the virtual photons 
exchanged between the incident beam and the target illuminate the nucleon constituents 
as the asymptotically free quarks of QCD. 
This is the regime of small separations between quarks and gluons, which is best tested 
with  short wavelength electromagnetic probes.  But the processes that polarized DIS explores 
are dominated by leading order terms in the expansion of $g_1$ that are related to the 
non-interacting partonic description of QCD.  
For a full understanding, the interactions 
among nucleon constituents must also be probed by looking into contributions beyond 
leading order, which grow in importance at lower energies.
%
%
Of particular interest are the so called `higher twist' terms, 
which measure correlations among partons (quarks and gluons), and offer
unique perspectives on the phenomenon of confinement inside nucleons and other hadrons.

Access to these subleading interactions is possible with transverse polarized inelastic scattering. In this configuration, the target spin is held orthogonal to the incident lepton spin, and the $g_2$ spin structure function dominates compared to $g_1$~\cite{Jaffe:1989xx,*Jaffe:1996zw}.  QCD matrix elements that arise from multi-parton correlations can be extracted directly from moments of $g_2$. However, this approach has been limited due to the technical challenges involved in aligning proton spins at 90$^\circ$ relative to the beam.
As we detail below, we have carried out a complete measurement of the
longitudinal and transverse inclusive SSF's $g_1$ and $g_2$ of the proton and
deuteron, with the goal of extracting matrix elements related to quark-gluon, $qg$, interactions.

The  Operator Product Expansion (OPE)~\cite{Ehrnsperger:1993hh,*Kodaira:1994ge} 
relates the moments of the SSF's to reduced quark and gluon matrix elements 
representing the multi-parton correlations that lead to confinement.
This perturbative approach to QCD is formulated in terms of the Cornwall-Norton (CN) 
moments~\cite{Melnitchouk:2005zr}:
\begin{eqnarray}
\Gamma_{1,2}^{(n)}(Q^2) = \int_0^1 dx~ x^{n-1} g_{1,2}(x,Q^2)
\label{eq:MOMENTS}
\end{eqnarray}
where 
$x= Q^2/2M\nu$ is the Bjorken scaling variable, 
$Q^2$ is the four-momentum transfer squared of the scattering process,
$M$ the nucleon mass, and $\nu$ the energy transfer.
Conventionally, the index $n=1$ is not displayed.

In the OPE,  the CN moments are expanded in a power series in $1/Q^{(\tau-2)}$ of increasing 
`twist' $\tau$, which
is defined as the mass dimension minus the spin $n$ of the corresponding QCD 
operator~\cite{Ehrnsperger:1993hh}. 
The leading twist-2 terms map to the successful predictions of the parton model,  
such as the $Q^2$-independence of the structure functions up to logarithmic corrections.
Higher twist (HT) corrections arise from the non-perturbative multiparton interactions,  
whose contributions at low energy increase as 
$1/Q^\tau$, reflecting confinement. Specifically, twist-3 represents 
$qgq$ correlations,
 so a non-zero result at a given $Q^2$  for the term representing twist-3 signals a
 departure from the non-interacting partonic regime.

Studies of higher twist effects~\cite{Meziani:2004ne,*Slifer:2008re,Ji:1997gs,Simula:2001iy,*Osipenko:2005nx,Prok:2008ev,Solvigno:2008hk} have typically focused on 
the CN moments 
$I(Q^2) = \int_0^1 dx~ x^2\left(2 g_1 + 3 g_2\right)$ and $\Gamma_1(Q^2)$,
in order to extract the twist-3  and twist-4  matrix  
elements.
However, as recently stressed in~\cite{Melnitchouk:2005zr,Dong:2008zz},
this approach is only appropriate when terms of purely kinematic origin, 
known
as `target mass corrections'
(TMC)~\cite{Nachtmann:1973mr,Wandzura:1977ce,Georgi:1976ve,Matsuda:1979ad,Sidorov:2006fi,Accardi:2008pc} 
for their connection to the finite nucleon mass,
can be neglected.
These terms, 
of order $\mathcal{O} (M^2/Q^2)$,
are formally related to twist-2 operators and must be
cleanly separated from the desired dynamical higher twists.    
$I(Q^2)$ is often labeled `$d_2$'
in the literature, although, as discussed below,
it is not equivalent to the HT matrix element indicated by the same symbol.

As emphasized in~\cite{Dong:2008zz}, dynamical HT can be extracted
to order $\mathcal{O}(M^8/Q^8)$  
by using
Nachtmann moments~\cite{Matsuda:1979ad,Piccione:1997zh}, which
depend on the scaling variable
$\xi= 2x/(1+\sqrt{1+(2xM)^2/Q^2})$. 
This variable extends scaling as $x$ approaches 1, for data measured at low $Q^2$. 
The Nachtmann moments are defined as
\begin{eqnarray}
\label{eq:NACMOMENTS}
\nonumber
M_1^n(Q^2)&\equiv& \frac{1}{2}\tilde{a}_{n-1}
\equiv \frac{1}{2}a_nE_1^n=
\int_0^1 \frac{dx}{x^2}\xi^{n+1}\\
\Bigl[
\Bigl\{\frac{x}{\xi}&-&\frac{n^2}{(n+2)^2}\frac{M^2 x\xi}{Q^2}  \Bigr\} g_1
-
\frac{4n}{n+2}\frac{M^2 x^2}{Q^2}g_2
\Bigr]~~~~~\\
\label{eq:NACMOMENTS2}
\nonumber
M_2^n(Q^2)&\equiv&\frac{1}{2} \tilde{d}_{n-1}
\equiv \frac{1}{2}d_nE_2^n=
\int_0^1 \frac{dx}{x^2}\xi^{n+1}\\
&&\Bigl[
\frac{x}{\xi} g_1
+
  \Bigl\{\frac{n}{n-1}\frac{x^2}{\xi^2}-
  \frac{n}{n+1}\frac{M^2x^2}{Q^2}\Bigr\}g_2
\Bigr]~~~~~
\end{eqnarray}
Here, the $a_n$ ($d_n$) represent twist-2 (-3) matrix elements, while
the $E_2^n$ are the corresponding Wilson coefficients, which contain logarithmic QCD corrections.
For convenience, these corrections are absorbed into the definition of the effective matrix 
elements $\tilde{a}_n$ and $\tilde{d}_{n}$. The index runs over $n=1,3,\ldots$ for $M_1$, and $n=3,5,\ldots$ for $M_2$.
When $Q^2{\gg}M^2$, the Nachtmann moments simplify:
\begin{eqnarray}
M_1^1(Q^2) \to  \Gamma_1(Q^2), ~~~~2 M_2^3(Q^2) \to I(Q^2)
\end{eqnarray}
Consequently, deviations  from unity of the ratio of Nachtmann to CN moments give a quantitative
measure of the TMCs at finite $Q^2$:
\begin{eqnarray}
R_1(Q^2)=\frac{M_1^1(Q^2)}{\Gamma_1(Q^2)}, ~~~~R_2(Q^2) = \frac{2M_2^3(Q^2)}{I(Q^2)}
\end{eqnarray}

The leading twist-2 contribution to $g_2$
can be determined
from $g_1$ via the Wandzura-Wilczek relation~\cite{Wandzura:1977qf}:
\begin{eqnarray}
g_2^\textrm{ww}(x,Q^2) \equiv -g_1(x,Q^2) + \int_x^1 {y^{-1}}{g_1(y,Q^2)} dy
\end{eqnarray}
So any twist-3  effects arise entirely from ${\overline g}_2 = g_2 -g_2^\textrm{ww}$.

\begin{center}
\begin{table}
\begin{tabular}{lcccc}
 &$\Gamma_1^p$ & $\Gamma_1^d$   & $\tilde{d}_2^p$  & $\tilde{d}_2^d$  \\
\hline
$f$              & 4.9  & 5.1   & 4.9  & 5.1  \\
RC               & 3.2  & 4.0   & 9.5  & 4.5  \\
$F_1$            & 3.0  & 4.0   & 3.0  & 4.0  \\
$R$              & 0.9  & 1.8   & 1.2  & 2.8  \\
$P_bP_t$         & 1.6  & 1.6   & 5.2  & 3.1  \\
$Q^2$ dependence & 0.33 & 0.1   & 4.3  & 0.3  \\
\hline
Total            & 6.8  & 8.0   & 13.0 & 8.9  \\
\end{tabular}
\caption{
Measured integral systematic uncertainties (in \%) arising from the target dilution factor $f$,
radiative corrections RC, the $F_1$ and $R$ data fits, beam and target polarizations
$P_b$ and $P_t$, and the evolution of our data fit to constant $Q^2$.
\label{tab:syst}
}
\end{table}
\end{center}

\begin{figure}[t!]
\includegraphics[angle=0,width=0.5\textwidth]{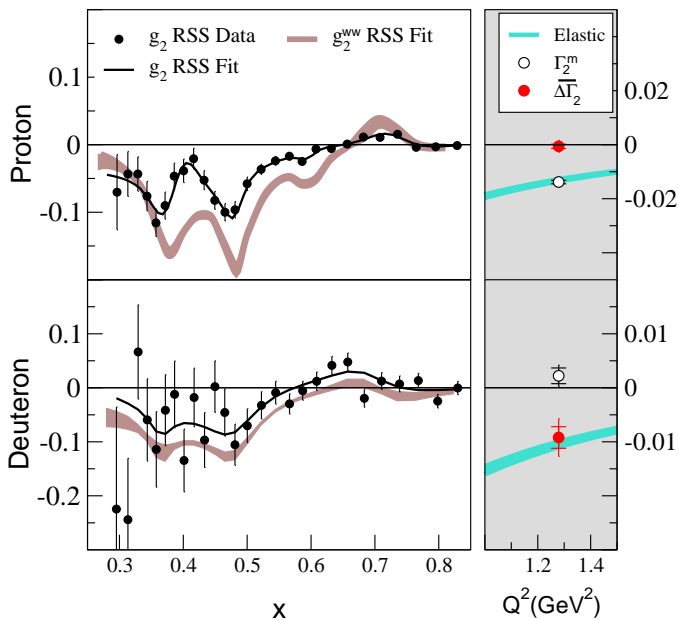} 
\caption{\label{fig:G2} 
{\bf Left-top:} 
RSS proton $g_2$ data~\cite{Wesselmann:2006mw}, along
with the RSS fit~\cite{Wesselmann:2006mw}.
The shaded curve is $g_2^\textrm{ww}$ evaluated from the RSS fit.
{\bf Left-bottom:} RSS deuteron $g_2$ data, and similar curves as above.
{\bf Right-top:} 
The open circle is the measured RSS proton $\Gamma_2^m(Q^2)$ data.
The full circle is $\Delta\overline\Gamma_2(Q^2)$.
The inner (outer)
error bars represent statistical (total) uncertainty.
The shaded curve is the elastic contribution to $\Gamma_2(Q^2)$.
{\bf Right-bottom:} Same as above, but for deuteron.
}
\end{figure}

Experiment E01-006 was conducted in Hall C of the
Thomas Jefferson National Accelerator Facility by the
Resonance Spin Structure (RSS) collaboration. 
We  measured the parallel and perpendicular 
double spin asymmetries $A_\parallel$ and $A_\perp$ 
in the scattering of 100 nA polarized
5.755 GeV electrons on polarized protons and deuterons.
Scattered electrons were detected 
at an angle of 13.15$^\circ$ using the Hall C High Momentum Spectrometer.
The kinematic coverage in invariant mass was $1.090 < W < 1.910$ GeV, corresponding
to $x_\mathrm{0}=0.316 < x < x_\mathrm{max}=0.823$,
at an average four-momentum transfer of $\langle Q^2 \rangle = 1.28\pm 0.21$ GeV$^2$.
Systematic uncertainties are detailed in Table~\ref{tab:syst}, with 
more details in~\cite{Wesselmann:2006mw} and~\cite{Jones:2006kf}.


The SSFs were extracted from the measured asymmetries using a fit to the 
ratio of longitudinal to transverse cross sections $R$~\cite{Bosted:2007xd,*Christy:2007ve} and the unpolarized structure function $F_1$~\cite{Bosted:2007xd,Christy:2007ve}, based primarily on data measured previously in Hall C.
The standard formulas and
procedure used to obtain 
$A_\parallel, A_\perp$ and the
SSFs from the data are detailed in~\cite{Abe:1998wq} and~\cite{Wesselmann:2006mw}.
Simultaneous determination of both SSFs 
allowed us to evaluate the 
moments 
without model input for $g_2$, as was  necessary in some previous analyses~\cite{Simula:2001iy,*Osipenko:2005nx}. 

The moments reported herein are evaluated at $\langle Q^2\rangle=1.28$ GeV$^2$, and have been decomposed 
%
%
into contributions from the measured resonance region $x_0 < x <x_{thr}$ 
(labeled with `m'), the well known $x = 1$
elastic (`el') contribution, and the unmeasured (`u') portion below
$x_0$. 
We note that the  small difference between our experimental $x_\mathrm{max}$ and
the nucleon inelastic threshold $x_\mathrm{thr}$ has negligible impact on the integrated results.
We have used
fits~\cite{Wesselmann:2006mw} to our data to evaluate the moments in the measured region.
%
%
The nucleon elastic contribution was calculated using the form factor
parameterizations of~\cite{Bradford:2006yz,*Arrington:2007ux}.  
Relative uncertainties of 5, 1, 14, and 2.5\% have been assumed for the electromagnetic form factors $G_{E,M}^P$ and $G_{E,M}^N$ respectively.
For the deuteron, $x=1$ represents quasielastic scattering, the strength of which we 
estimate by combining the nucleon elastic contributions using the $D$-state correction discussed below.
The deuteron nuclear elastic contribution is negligible here.
%
%
We have also evaluated the neutron and non-singlet (proton - neutron)  moments
by using the relation~\cite{CiofidegliAtti:1996uc} 
$\Gamma^n=\Gamma^\Sigma - \Gamma^p$, where the singlet $\Gamma^\Sigma = \Gamma^d/\gamma_D$, and
$\gamma_D  \simeq 0.926\pm0.016$~\cite{Rondon:1999da} is the $D$-state correction to the deuteron
wave function. The uncertainty arising from this approach is
estimated~\cite{Kulagin:2007ph,*Kahn:2008nq} to be $\mathcal{O}(1\%)$. 
The singlet and non-singlet results assume negligible 
heavy quark contributions.
\begin{table*}[]
\begin{tabular}{|c@{ }|l@{}|l@{~~}l@{~~}l|l@{~~}l@{~}l||l@{~~}l@{~~}|l@{~~}l@{~}|}
\hline
&&\multicolumn{3}{|c|}{Proton} &\multicolumn{3}{|c||}{Deuteron} &\multicolumn{2}{|c|}{Neutron}&\multicolumn{2}{|c|}{Non-Singlet}  
\\
$x$-range&
&Value & Stat & Syst & Value& Stat & Syst & Value& Total & Value& Total\\
\hline
&$M_{1}^{1}$ &
     0.0676&     $\cdots$&     0.0069&
     0.0274&     $\cdots$&     0.0104&
    -0.0381&     0.0132&
     0.1057&     0.0178
\\
$0<x< x_{0}$
&$\Gamma_1$ &
     0.0683&     $\cdots$&     0.0069&
     0.0280&     $\cdots$&     0.0099&
    -0.0381&     0.0127&
     0.1065&     0.0176
\\
(unmeasured)
&$\tilde{d_2}$ &
        0.0&     $\cdots$&     0.0008&
        0.0&     $\cdots$&     0.0013&
        0.0&     0.0016&
        0.0&     0.0021
\\
&$I$ &
        0.0&     $\cdots$&     0.0008&
        0.0&     $\cdots$&     0.0013&
        0.0&     0.0016&
        0.0&     0.0021
\\
\hline
&$M_{1}^{1}$ &
     0.0330&     0.0005&     0.0022&
     0.0290&     0.0010&     0.0023&
    -0.0016&     0.0036&
     0.0346&     0.0054
\\
$x_{0}\le x<x_{thr}$
&$\Gamma_1$ &
     0.0351&     0.0005&     0.0024&
     0.0315&     0.0011&     0.0025&
    -0.0010&     0.0039&
     0.0361&     0.0058
\\
(measured)
&$\tilde{d_2}$ &
     0.0037&     0.0004&     0.0005&
     0.0048&     0.0008&     0.0004&
     0.0015&     0.0012&
     0.0022&     0.0016
\\
&$I$ &
     0.0057&     0.0006&     0.0007&
     0.0082&     0.0013&     0.0007&
     0.0031&     0.0019&
     0.0026&     0.0026
\\
\hline
&$M_{1}^{1}$ &
     0.0287&     $\cdots$&     0.0006&
     0.0338&     $\cdots$&     0.0012&
     0.0078&     0.0005&
     0.0208&     0.0008
\\
$x=1$
&$\Gamma_1$ &
     0.0351&     $\cdots$&     0.0010&
     0.0373&     $\cdots$&     0.0015&
     0.0051&     0.0007&
     0.0300&     0.0012
\\
(elastic)
&$\tilde{d_2}$ &
     0.0067&     $\cdots$&     0.0008&
    -0.0021&     $\cdots$&     0.0010&
    -0.0090&     0.0007&
     0.0157&     0.0011
\\
&$I$ &
     0.0306&     $\cdots$&     0.0025&
     0.0088&     $\cdots$&     0.0029&
    -0.0212&     0.0019&
     0.0518&     0.0031
\\
\hline
&$M_{1}^{1}$ &
     0.1293&     0.0005&     0.0073&
     0.0902&     0.0010&     0.0108&
    -0.0318&     0.0137&
     0.1611&     0.0187
\\
$0\le x\le 1$
&$\Gamma_1$ &
     0.1385&     0.0005&     0.0074&
     0.0968&     0.0011&     0.0103&
    -0.0340&     0.0134&
     0.1725&     0.0185
\\
(Total)
&$\tilde{d_2}$ &
     0.0104&     0.0004&     0.0013&
     0.0027&     0.0008&     0.0017&
    -0.0075&     0.0021&
     0.0179&     0.0029
\\
&$I$ &
     0.0364&     0.0006&     0.0027&
     0.0170&     0.0013&     0.0032&
    -0.0180&     0.0031&
     0.0544&     0.0045
\\
\hline
$0\le x\le 1$&$R_1$ &
      0.933&\multicolumn{2}{c|}{      0.003}&
      0.932&\multicolumn{2}{c||}{      0.012}&
      0.936&      0.037&
      0.934&      0.008
\\
&$R_2$ &
      0.286&\multicolumn{2}{c|}{      0.014}&
      0.159&\multicolumn{2}{c||}{      0.078}&
      0.415&      0.045&
      0.329&      0.025
\\
\hline
\end{tabular}
\caption{
\label{table:d2}
RSS Moments evaluated at $\langle Q^2\rangle =1.28$ GeV$^2$. 
The ratio $R_1=M_1^1/\Gamma_1$ and $R_2=\tilde{d_2}/I$.
}
\end{table*}

Table~\ref{table:d2} provides numerical values for the moments.
The ratio $R_2$ differs significantly from unity, indicating large unwanted twist-2
kinematical contributions to $I(Q^2)$.
The full results for the   matrix element $\tilde d_{2}$ indicate clear 
twist-3 at more than 6$\sigma$ for the proton and 3$\sigma$ for the neutron.
These non-vanishing results unambiguously indicate the presence of 
$qgq$ correlations.
Their
magnitudes can be used in comparing with nucleon models.


The values of $I(Q^2)$ and $\tilde{d_2}$ for $x < x_0$
were estimated to be zero, with a systematic error described below.
In this unmeasured region,  $\xi \approx x$ and the CN and Nachtmann truncated moments converge,
so there is little difference between  $I(Q^2)$ and $\tilde d_2$.
$I(Q^2)$ is expected to be small in
this unmeasured region, because of the decreasing importance of higher twists at low $x$, 
and the strong suppression due to the $x^2$ weighting of the integral.
This assumed low $x$ behavior is supported by our data.
%
%
%
Figure~\ref{fig:G2} (left)  displays the $x-$dependence of the measured $g_2$ structure function.
It is clear that $\overline{g}_2$ is nearly constant and
consistent with zero within errors for the proton, near the low end of our measured range $x_0< x \lesssim 0.37$.
The deuteron data shows a similar behavior,
different from zero only at the one $\sigma$ (statistical) level.
Neutron data~\cite{Kramer:2005qe} at 
$x\sim0.2$ and similar $Q^2$ are
also consistent with ${\overline g_2} \cong 0$.
%
%
%
%
%
We take the error on $\overline g_2$, evaluated at $x_0$, $\delta \overline g_2$, as a conservative
upper limit for the integrand of $I(Q^2)$ in the unmeasured low $x$ region.
For the proton, we determined this upper limit 
by assuming a constant extrapolation of the value of $\delta{\overline g}_2(x_0)$ down to $x=0$.
For the deuteron, we evaluated both a linear and constant extrapolation,
averaged both assumptions and added quadratically one-half their difference as `model' error to
the $\delta \overline g_2^d$ fit error to obtain the value in Table~\ref{table:d2}.
A divergence of $g_2$ as $x\to 0$ could invalidate this assumption.  
Such a possible divergence for 
$x \lesssim 0.001$ was discussed
in~\cite{Ivanov:1999gq}. Normalizing the low $x$ dependence of $\overline g_2$ given in~\cite{Ivanov:1999gq} 
to our $\delta \overline g_2$ estimates, we find 
the additional contribution to be negligible.


It is instructive to compare our twist-3 results to previous measurements. 
SLAC E155~\cite{Anthony:2002hy}  reported an evaluation of 
$I(Q^2)$ at $\langle Q^2\rangle=5$ GeV$^2$.
We have corrected those results for TMCs~\cite{Dong:2008zz}, to obtain
$0.0028\pm0.0015$ and $0.0072\pm0.0044$ for the proton and neutron, respectively.  
For direct comparison, we  performed a 
pQCD evolution~\cite{Shuryak:1981pi,*Ji:1990br} 
from RSS to the SLAC kinematics. At LO, we find  $\tilde d_2^p=0.0021\pm0.0006$ and
$\tilde d_2^n=0.0031\pm0.0038$, which
are consistent with the E155 results.
The elastic contribution~\cite{Bradford:2006yz,*Arrington:2007ux} at these kinematics is smaller than the uncertainties and is not included in the results of this paragraph.  
NLO corrections~\cite{Ji:2000ny,*Belitsky:2000pb}  to our  
data 
have been calculated and 
will be discussed in a future publication.

Twist-3 effects also manifest in the first moment of $g_2$.
The Burkhardt-Cottingham (BC) 
sum rule~\cite{Burkhardt:1970ti} predicts that $\Gamma_2$ vanishes for all $Q^2$.
%
%
This sum rule can be derived from the unsubtracted dispersion relation for the 
virtual-virtual Compton scattering  amplitudes, 
in an analogous 
fashion 
to the more famous
GDH and Bjorken sum rules.
It provides a remarkably clean tool to investigate nucleon structure, since it is free
from both QCD radiative corrections and TMCs~\cite{Kodaira:1994gh}.
%
Initial measurements by the E155 collaboration~\cite{Anthony:2002hy}
found a 2$\sigma$ inconsistency with the proton BC sum rule at large $Q^2$,
while the same group found the deuteron sum rule to hold.
Later 
measurements~\cite{Amarian:2003jy,Slifer:2008re} at lower momentum transfer
found agreement with  both the
neutron and $^3$He BC sum rules. 

The leading twist $g_2^\textrm{ww}$ satisfies the BC sum rule exactly, so any violation must arise
from higher twist effects.  
For convenience we split the full integral as:
\begin{eqnarray}
\nonumber
\Gamma_2 &=& \Gamma_2^\textrm{ww} + \overline \Gamma_2 + \Gamma_2^{el}\\
         &=& \overline \Gamma_2^u + \overline \Gamma_2^m + \Gamma_2^{el}
\end{eqnarray}
where 
the overbar signifies removal of the leading twist contribution, and
we have 
made use 
of the fact that $\Gamma_2^\textrm{ww} \equiv 0$ by definition.  The elastic contribution is well known, as displayed in Fig.~\ref{fig:G2}, and we can evaluate $\overline{\Gamma}_2^m$ directly from our own data.  
$\overline{\Gamma}_2^u$ is pure higher twist, and cannot be directly determined, due to the lack of low $x$ $g_2$ data.  However, 
the difference, 
\begin{eqnarray}
\Delta\overline\Gamma_2 
&\equiv& \Gamma_2-\overline \Gamma_2^u 
\end{eqnarray}
depends only on measured data.  A significant non-zero result for
this quantity
would indicate 
a higher twist contribution to the integral in the region $x<x_0$, assuming of course, that the BC sum rule holds.
%
We find that 
the proton results are consistent with vanishing HT effects at low $x$.  On the other hand, the singlet results indicate the need for significant HT contributions to $\Gamma_2$ at low $x$, if the BC sum rule is to be satisfied.  
This could indicate an isospin dependence of the BC sum rule, or alternatively, a 
modification to the sum rule due to nuclear effects.

\begin{table*}[]
\begin{tabular}{|c@{}|r@{~~}c@{~~}c|r@{~~}c@{~}c|r@{~~}c@{~~}|r@{~~}c@{~}|}
\hline
&\multicolumn{3}{|c|}{Proton} &\multicolumn{3}{|c|}{Deuteron} &\multicolumn{2}{|c|}{Neutron}&\multicolumn{2}{|c|}{Non-Singlet}  
\\
%
&Value & Stat & Syst & Value& Stat & Syst & Value& Total & Value& Total\\
\hline
$\Gamma_2^{el}$ &
    -0.0132&     $\cdots$&     0.0004&
    -0.0219&     $\cdots$&     0.0009&
    -0.0105&     0.0005&
    -0.0027&     0.0007
\\
$\Gamma_2^m$ &
    -0.0138&     0.0007&     0.0009&
    -0.0107&     0.0012&     0.0009&
     0.0022&     0.0020&
    -0.0160&     0.0028
\\
$\overline{\Gamma}_2^m$ &
     0.0126&     0.0008&     0.0020&
     0.0129&     0.0016&     0.0018&
     0.0013&     0.0034&
     0.0113&     0.0051
\\
\hline
$\Delta\overline\Gamma_2$ &
    -0.0006&     0.0008&     0.0021&
    -0.0090&     0.0016&     0.0020&
    -0.0092&     0.0035&
     0.0086&     0.0051
\\
\hline
\end{tabular}
\caption{
\label{table:GAM2}
The higher twist $\Delta\overline\Gamma_2=\Gamma_2 - \overline \Gamma_2^u$ evaluated at $\langle Q^2\rangle =1.28$ GeV$^2$. 
}
\end{table*}
Nachtmann moments are also useful for obtaining leading twist matrix
elements 
free of 
TMCs.
The $a_0$ non-singlet matrix
element corresponds to the lowest moment $\Gamma_1^{p-n}$ of the isovector
nucleon $g_1$, which is related to the 
nucleon axial charge $g_A$ via 
the Bjorken sum rule~\cite{Bjorken:1966jh}.  
This relation is a direct consequence of QCD, and experimental tests~\cite{Anthony:2000fn,Airapetian:2007mh}
of the sum rule 
have played a critical role 
in confirming QCD as the correct theory of the strong interaction.  
The target mass corrections to $\Gamma_1$ given by the ratio $R_1$
are small at our kinematics, due to a combination of dominant contributions from the
region below our lowest measured $x_\mathrm{0}$, and the nearly
identical values of $x$ and $\xi$ in that region. 
It is interesting to note that $R_1\approx 93\%$ for all targets, when the
elastic contribution is included.  
The $M_1^1$ non-singlet result including the elastic part ($0.161\pm 0.019$) is in excellent agreement
with theory: Ref.~\cite{Eidelman:2004wy} predicts 0.212, which reduces to 0.155  after
application of the next to next-to-leading order 
corrections~\cite{Kataev:1994gd}
appropriate at our $Q^2$. 
For the $g_1$ integrals, the contribution from the  $x<x_0$ region was calculated from Regge-inspired fits to SLAC
$g_1^p$ and $g_1^d$ data~\cite{Abe:1998wq,Anthony:2000fn} within a band $Q^2 = 1.3 \pm 0.3$
GeV$^2$.  
We find $g_1 = a x^b (1-x)^3(1+c/Q^2)$
with $a=0.392\pm 0.254$,  
$b=0.0676\pm 0.084$, 
$c=0.0636\pm 0.681$ for the proton,
and $a=1.778\pm 1.971$, $b=0.739\pm 0.407$, and
$c=-0.786\pm 0.0256$ for the deuteron.

In summary, we have measured the spin structure function moments
of the proton, deuteron and neutron in the resonance region.
The kinematic weighting of higher moments suppresses contributions from the low $x$ region, 
thus minimizing the systematic uncertainty of this unmeasured piece of the integral.
Our analysis indicates a large TMC contamination of the CN moment $I(Q^2)$, 
which has often been used to determine the magnitude of higher twist effects.
The alternate Nachtmann moment $\tilde{d_2}$ minimizes TMCs and allows a clean extraction of
the twist-3 reduced matrix element.  
The previously undetermined $\tilde{d_2}$ provides
unambiguous evidence of dynamical twist-3 effects, which arise from quark-gluon interactions.
The first moment of $g_2$ indicates an isospin dependence of HT effects in the region of small
Bjorken scaling variable, assuming that the BC sum rule holds.  
After nearly thirty years of spin-dependent investigations, the 
$g_2^p$ structure function remains largely unknown.  But the data reported here and 
future analyses~\cite{G2} will shed light on this fundamental nucleon observable.

We would like to thank the Hall~C technical staff and
the accelerator operators for their efforts and dedication.
This work was supported by
the Department of Energy,
the National Science Foundation,
the Schweizerische Nationalfonds,
and by the
Institute of Nuclear and Particle Physics of the University of Virginia.
The Southeastern Universities
Research Association  operates the Thomas Jefferson National Accelerator
Facility for the DOE under contract DE-AC05-84ER40150, mod. \# 175.

\bibliography{rssm}

\end{document}